# The Effect of Planetary Rotation Period on Clouds in a Global Climate Model with a Bin Microphysics Scheme

Huanzhou Yang[1], Eric T. Wolf[2,3,4,5,6], Cheng-Cheng Liu (刘成诚)[2], Yunqian Zhu[7,8], Owen B. Toon[2,3], and Dorian S. Abbot[1]
[1] Department of the Geophysical Sciences, University of Chicago, USA; jeffyang35@gmail.com
[2] Laboratory for Atmospheric and Space Physics, University of Colorado Boulder, USA
[3] Department of Atmospheric and Oceanic Sciences, University of Colorado Boulder, USA
[4] NASA GSFC Sellers Exoplanet Environments Collaboration, USA
[5] NASA NExSS Virtual Planetary Laboratory, USA
[6] Blue Marble Space Institute of Science, USA
[7] Cooperative Institute for Research in Environmental Sciences, University of Colorado Boulder, USA
[8] Chemical Sciences Laboratory, National Oceanic and Atmospheric Administration, USA

## Abstract

Clouds are the largest source of uncertainty in climate simulations. For exoplanets, cloud simulation is particularly challenging because of the lack of observational data to tune parameterized cloud models. Here we apply Community Aerosol and Radiation Model for Atmospheres (CARMA), a size-resolved bin cloud microphysics model, to the atmospheric global climate model Community Atmosphere Model (CAM6) and simulate exoplanets with a range of planetary rotation rates. CARMA produces fewer liquid clouds than the native CAM6 parameterized cloud microphysics scheme (Morrison–Gettelman two-moment microphysics, MG), more ice clouds, and a significantly different ice cloud size distribution. Overall, this leads to a decrease in the magnitude of the net CRE by 4–10 W m$^{-2}$, which is unlikely to change the determination of habitability from a climate perspective in most cases. The difference in ice cloud size distribution is likely to strongly affect transmission spectral retrievals. Our work confirms that the MG parameterized cloud microphysics scheme can produce reasonable climate simulation when extrapolated to some exoplanet contexts and highlights the value of resolved cloud microphysics for evaluating parameterized schemes and for interpreting observations.

## 1. Introduction

A grand challenge in astronomy is determining the possibility for habitable conditions and the presence of life on exoplanets. So far this search has focused mostly on relatively cold stars and close-in planets using transit spectroscopy with missions such as Hubble Space Telescope, TESS, and JWST (J. Mather 2003; D. K. Sing et al. 2011; G. R. Ricker et al. 2015), and it has not yielded strong evidence for habitable nor inhabited exoplanets. But the lack of definitive evidence to date may be in large part due to the limitations of our current observing capabilities. The Habitable Worlds Observatory (HWO; National Academies of Sciences, Engineering, and Medicine 2021), expected to be launched in the 2040s, will be optimized for characterizing Earth-like planets orbiting Sun-like stars using direct imaging and reflected light spectroscopy. Detailed climate modeling of potentially habitable exoplanets is necessary for identifying observational targets, interpreting observations, and planning HWO. Clouds will have a strong effect on reflected light spectra (M. S. Marley et al. 1999; M. C. Turnbull et al. 2006; Y. Kawashima & S. Rugheimer 2019) and thus need to be considered with increased rigor if the community is to have confidence in our interpretations.

Clouds are the main source of uncertainty in climate modeling. This is because they are significantly subgrid scale (∼$10^2$–$10^3$ m) relative to the grid boxes (∼$10^5$ m) of global climate models (GCMs), the standard tool for simulating the global climate of a planet. As a result, they are parameterized. Therefore, a function must be found that predicts the average subgrid-scale behavior of clouds based on grid-scale quantities such as temperature, humidity, and winds, combined with adjustable parameters controlling things such as convective timescales, cloud fractions, precipitation autoconversion, and assumptions on particle sizes (E. T. Wolf et al. 2022). These tunable parameters are typically benchmarked against modern Earth conditions, which differ among GCMs, and these differences are accentuated when they are applied to exoplanet contexts where their parameterizations were not tuned for (J. Yang et al. 2019; D. E. Sergeev et al. 2022). This has led to the deployment of computationally expensive cloud-permitting or cloud-resolving models in exoplanet simulation, in which the model grid scale is reduced to nearly the cloud scale (D. E. Sergeev et al. 2020, 2024; M. Lefèvre et al. 2021; J. Liu et al. 2023; J. Yang et al. 2023).

The process-simulation scale mismatch is even more dire for cloud particles, which are on the scale of ∼$10^{-6}$–$10^{-3}$ m and therefore hopelessly unresolved even in cloud spatial resolving models. This is important because so-called cloud microphysics govern the cloud effect on radiation and therefore strongly influence both climate and observations (T. D. Komacek et al. 2020). Cloud microphysics schemes vary across a range of complexities, with the simplest assuming a constant effective cloud particle radius or number of cloud particles (B. A. Boville et al. 2006; J. Edwards et al. 2007; F. Hourdin et al. 2013) and more sophisticated multimoment schemes depending on temperature, saturation ratio, and aerosols (H. Morrison & A. Gettelman 2008; A. Gettelman et al. 2015). Even the most sophisticated cloud microphysics schemes have been tuned to Earth, and their application to exoplanets represents an extrapolation.

To begin addressing the effect of cloud microphysics on exoplanets, the bin-resolved, process-based cloud microphysical model (H. Yang et al. 2024) applied the Community Aerosol and Radiation Model for Atmospheres (CARMA; R. Turco et al. 1979; O. B. Toon et al. 1988) in one vertical dimension. CARMA explicitly simulates cloud microphysical processes including nucleation, growth, coagulation, coalescence, and evaporation from first principles and therefore should yield more dependable predictions when used to extrapolate to exoplanets. H. Yang et al. (2024) systematically varied planetary parameters and found that while macrophysical features such as temperature and moisture dominate the cloud radiative effect (CRE), planetary parameters such as stellar flux and atmospheric surface pressure can significantly affect the size distribution of cloud particles. By feeding the 1D CARMA output into the Planetary Spectrum Generator (G. L. Villanueva et al. 2022), they found that clouds should enhance HWO observations because they provide extra reflection to increase the number of photons to be detected (H. Yang et al. 2025). This 1D effort only modeled the average state of a planet, neglecting spatially heterogeneous differential cloud formation mechanisms involving 3D global circulation.

In this paper, we will study the effect of bin-resolved cloud microphysics on cloud behavior and climate in a 3D GCM in exoplanet-relevant simulations. We will vary the planetary rotation period because it is an inherently 3D parameter known to strongly influence 3D circulation. The impact of rotation period on exoplanet climate has previously been studied using GCMs without clouds or with parameterized clouds and cloud microphysics (J. Yang et al. 2014; Y. Kaspi & A. P. Showman 2015; S. Faulk et al. 2017; T. D. Komacek & D. S. Abbot 2019), providing important context for interpreting our simulations.

This paper is organized as follows. We introduce the models that we use and the experimental setup in Section 2. Section 3 presents results, and Section 4 discusses the context and limits of this work. Section 5 summarizes our main findings.

## 2. Methods

### 2.1. Model Description

We use version 6 of the Community Atmosphere Model (CAM6), which is a component of the 3D GCM CESM2.1.1 (G. Danabasoglu et al. 2020). This is a developing version, and the first release of CESM2 with CARMA will be in CESM2.2.2. To simulate planets with different rotation periods, we apply a subset of code modifications from ExoCAM (E. T. Wolf et al. 2022), which is based on CESM1/CAM4, in order to facilitate generalization of the diurnal cycle to any arbitrary rotation period. CAM6 simulates macrophysical aspects of clouds, such as cloud fraction, with the Cloud Layers Unified by Binormals (CLUBB; J.-C. Golaz et al. 2002; V. E. Larson et al. 2012; P. A. Bogenschutz et al. 2013). The CLUBB simulations include moist turbulence in the boundary layer and shallow cumulus convection. Deep convection is simulated by the Zhang–McFarlance (G. J. Zhang & N. A. McFarlane 1995) scheme. The radiative transfer model is the Rapid Radiative Transfer Model for Global Climate Models (RRTMG), the GCM-coupled version of RRTM (E. J. Mlawer et al. 1997). When CARMA cloud is used as the cloud microphyiscs module, we calculate the radiative properties from CARMA and pass to RRTMG. Though usually applied for Earth-like simulations, RRTMG is compatible with exoplanets with similar CO_2 levels to modern Earth and stellar spectrum similar to the Sun.

Morrison–Gettelman two-moment microphysics (hereafter MG; H. Morrison & A. Gettelman 2008) is the native cloud microphysics model in CAM6. MG tracks cloud mass and number and calculates the effective radius for both ice and liquid clouds. Crucially, MG parameterizes cloud size using a $\Gamma$-distribution that depends on the temperature and humidity instead of resolving cloud particle concentrations in bins. We also perform simulations with a version of CAM6 that includes the CARMA cloud microphysics model in each atmospheric column. Microphysical species, including aerosols and cloud particles, are represented by “groups” in CARMA. There are multiple atmospheric aerosol species, including dust, sea salt, and sulfate, that can be activated and serve as as cloud condensation nuclei or ice nuclei. The cloud species include liquid droplet, in situ and detrained ice particles, and graupel, which are all represented by 48 size bins. Liquid particle nucleation, growth, and evaporation are implemented as in H. Yang et al. (2024), and analogous processes for ice particles are implemented as in C. Bardeen et al. (2013). A detailed description can be found in C.-C. Liu et al. (2026). For the atmospheric lower boundary condition we use Earth’s continental distribution, which provides CARMA with different types of surface aerosol fluxes for cloud nucleation, including dust, sea salt, and sulfate. CARMA is capable of modeling aerosols, which are dry particles like dust, sea salt, and other chemically formed species (S. Tilmes et al. 2024). However, to reduce computation time and focus on clouds, we simulate aerosols with CAM6’s native Modal Aerosol Module both in CARMA and MG simulations.

### 2.2. Experimental Setup

We run CAM6 at a horizontal resolution of $1\overset{\circ}{.}875 \times 2\overset{\circ}{.}5$ (latitude by longitude) with 56 vertical layers. We calculate the cloud water content and effective radius from mass bins. Unlike MG, CARMA does not distinguish between cloud particles and precipitation particles such as rain and snow. To be consistent with MG (H. Morrison & A. Gettelman 2008), we use maximum cloud radius cutoffs of 50 $\mu$m for liquid and 400 $\mu$m for ice. Larger particles are considered rain or snow. We run the model at four different rotation periods: 36.5, 2, 1, and 0.5 days, where we use “day” as a time unit representing a standard Earth day. For all simulations, we use modern Earth’s orbital configuration. The computational cost of CAM6-CARMA is $\sim$14000 cpu-hr per model year, which is 14 times that CAM6-MG ($\sim$1000 cpu-hr per model year). This makes CAM6-CARMA prohibitively expensive to run simulations from startup to new equilibrium, which can take several decades or longer depending on the assumed initial conditions and planetary parameters. As a result, we do not run CAM6-CARMA simulations to equilibrium. Instead, we use the following methodology. First we run CAM6-MG simulations with a mixed layer ocean to steady state. For this we use a uniform 50 m mixed layer depth and a time-invariant ocean heat transport derived from modern Earth's climate. We use the resulting seasonally varying sea surface temperature and sea ice distributions as boundary conditions for a 1 yr CAM6-CARMA simulation. This methodology allows us to isolate the differences in cloud behavior between CAM6-MG and CAM6-CARMA in a very similar climate state. Note that ocean heat transport is also sensitive to rotation period. For example, S. L. Olson et al. (2020) used ROCKE-3D with dynamic ocean and found that halving the rotation rate could yield a cooling of about 7 K relative to the nominal scenario of Earth. This difference is significant compared to the temperature differences in our work that consider only the atmosphere reaction to rotation period. The interaction of ocean and cloud effects as the rotation rate is changed should be considered in future work.

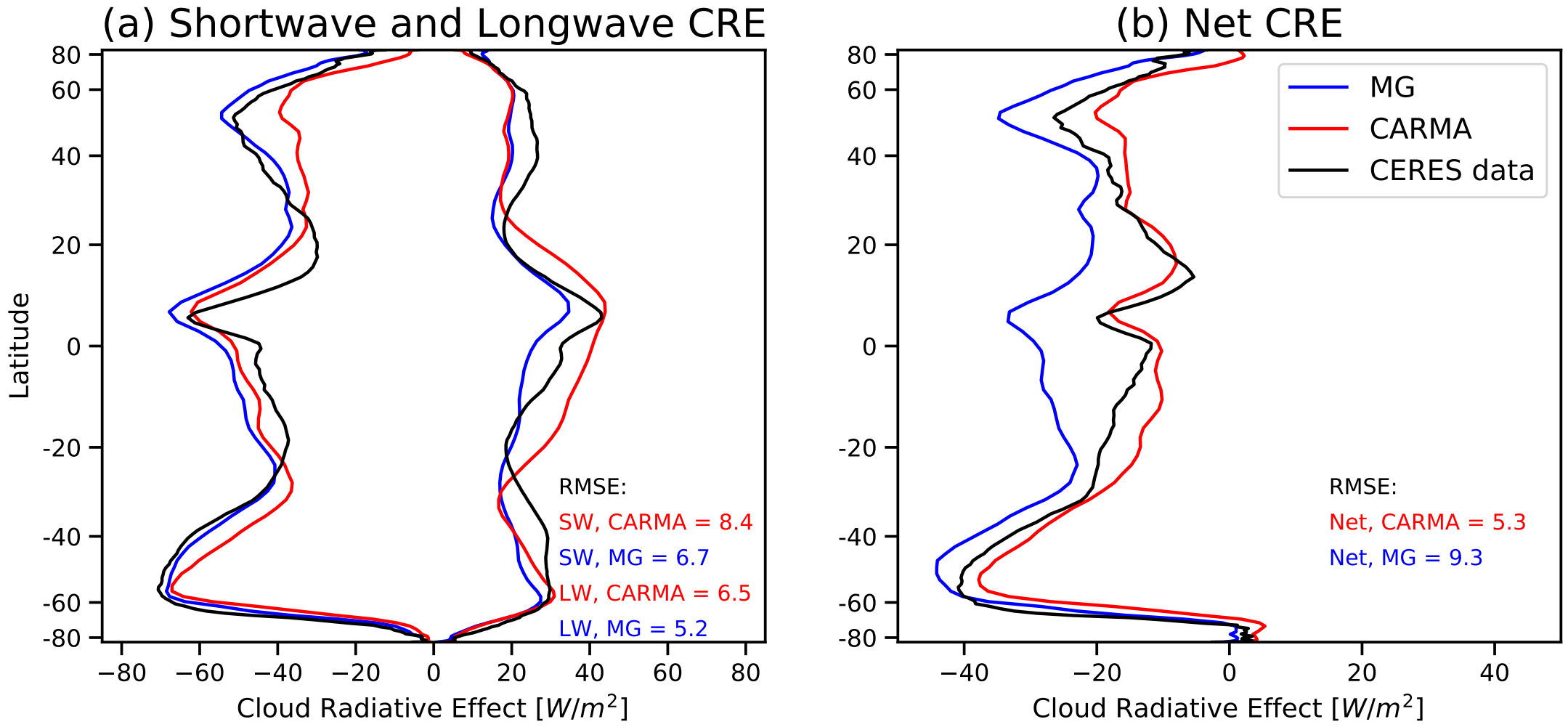

**Figure 1.** Shortwave and longwave CRE (panel (a)) and net CRE (panel (b)) for simulations with MG (blue) and CARMA (red) at Earth's rotation period as well as CERES observational data (black). The area-weighted root mean squared error (RMSE) of both models relative to CERES data is also shown.

## 3. Results

To validate CARMA, we compare the zonal mean CRE in CARMA and MG at Earth's rotation period with CERES satellite observations (Figure 1). More extensive validation of CARMA was performed by C.-C. Liu et al. (2026) and will appear in a forthcoming paper by the same author. Both MG and CARMA capture the overall latitudinal pattern of longwave, shortwave, and net CRE observed in data. Interestingly, while MG fits both the longwave and shortwave CRE observations somewhat better than CARMA, CARMA fits the net CRE much better than MG due to canceling errors. The most noticeable differences between MG and CARMA, which will become important as we change the rotation period below, are that CARMA has a more positive longwave CRE in the tropics and a more negative shortwave CRE in the mid-latitudes (Figure 1). It is important to note that MG is part of the standard CESM2 package and has been more extensively tuned to modern Earth than CARMA. Moreover, the risk of overfitting is likely to be lower in CARMA than MG because it more explicitly models cloud microphysical processes.

As a first pass at investigating how the differences between MG and CARMA clouds may effect habitable exoplanet climates, we vary the rotation period. We show global-mean values of important variables in Figure 2. The global-mean temperature decreases with increasing lengths for the rotation period (Figure 2(a)). This is mainly due to a decrease in low-latitude temperature associated with a decrease in the equator-to-pole temperature gradient (not shown), consistent with Y. Kaspi & A. P. Showman (2015). In both models increasing the rotation period makes the shortwave CRE more negative (Figure 2(d)), which is consistent with increases in liquid cloud water path (CWP; Figure 2(b)). Increases in liquid CWP and the magnitude of shortwave CRE are particularly pronounced in MG for rotation periods less than 2 days. Changes in longwave CRE with rotation period (Figure 2(e)) are smaller than changes in shortwave CRE, and the ice CWP is fairly constant and significantly larger in CARMA than MG (Figure 2(c)). The trend in longwave CRE flattens (MG) or reverses (CARMA) at a rotation period of 36.5 days. This is due both to a flattening or decrease in ice CWP and, as we will see later, a decrease in tropical cloud height. Regardless of rotation period, CARMA always has a less negative shortwave CRE (Figure 2(d)) and more positive longwave CRE (Figure 2(e)). As a result, the net CRE in MG is 4–10 W $m^{-2}$ more negative than that in CARMA (Figure 2(f)). The global-mean values also give us hints that cloud microphysics can have effects on CRE beyond simply changing the CWP. For example, at a rotation period of 0.5 day, MG has a lower water CWP (Figure 2(b)) but a stronger shortwave CRE (Figure 2(d)).

The shortwave CRE is dominated by liquid particles. In MG this is simply because the liquid CWP (Figure 2(b)) is 1 order of magnitude larger than the ice CWP (Figure 2(c)). In CARMA the liquid CWP (Figure 2(b)) is only a factor of 2 larger than the ice CWP (Figure 2(c)), but most of the additional ice particles have radii larger than 100 $\mu$m and interact weakly with shortwave radiation. We can therefore understand the effect of rotation period and cloud microphysics on shortwave CRE by studying the liquid CWP and the distribution of liquid droplet effective radius. At rotation periods of 0.5 and 1 day, there are two major regions that contribute to shortwave CRE: the equatorial region where convective clouds form in the ITCZ and the mid-latitudes where extratropical storms result in massive clouds (Figures 3 and 4). Between them is an arid and relatively cloudless subtropics where the Hadley cell downwells. As the rotation period increases, the Hadley cell expands, becoming nearly global at a rotation period of 36.5 days, such that the relatively cloudless subtropical region vanishes. For all rotation periods, CARMA has more liquid CWP than MG at low latitudes and less liquid CWP at latitudes higher than $\sim 40^\circ$ (Figure 4). Additionally, CARMA has a larger liquid water effective radius at all rotation periods, although the difference between the two models decreases as the rotation period increases (Figure 6). The optical depth of clouds can be approximated as $\mathrm{CWP}/(2R_{\mathrm{eff}})$. The smaller liquid CWP at latitudes higher than $\sim 40^\circ$ and the larger liquid water effective radius in CARMA both contribute to a less negative shortwave CRE in the mid-latitudes (Figure 3), which is the main reason that the global-mean shortwave CRE is less negative in CARMA than MG for all rotation periods (Figure 2(d)). In the tropics, the larger liquid CWP and larger liquid cloud effective radius in CARMA roughly cancel so that CARMA and MG have similar shortwave CRE (Figure 3).

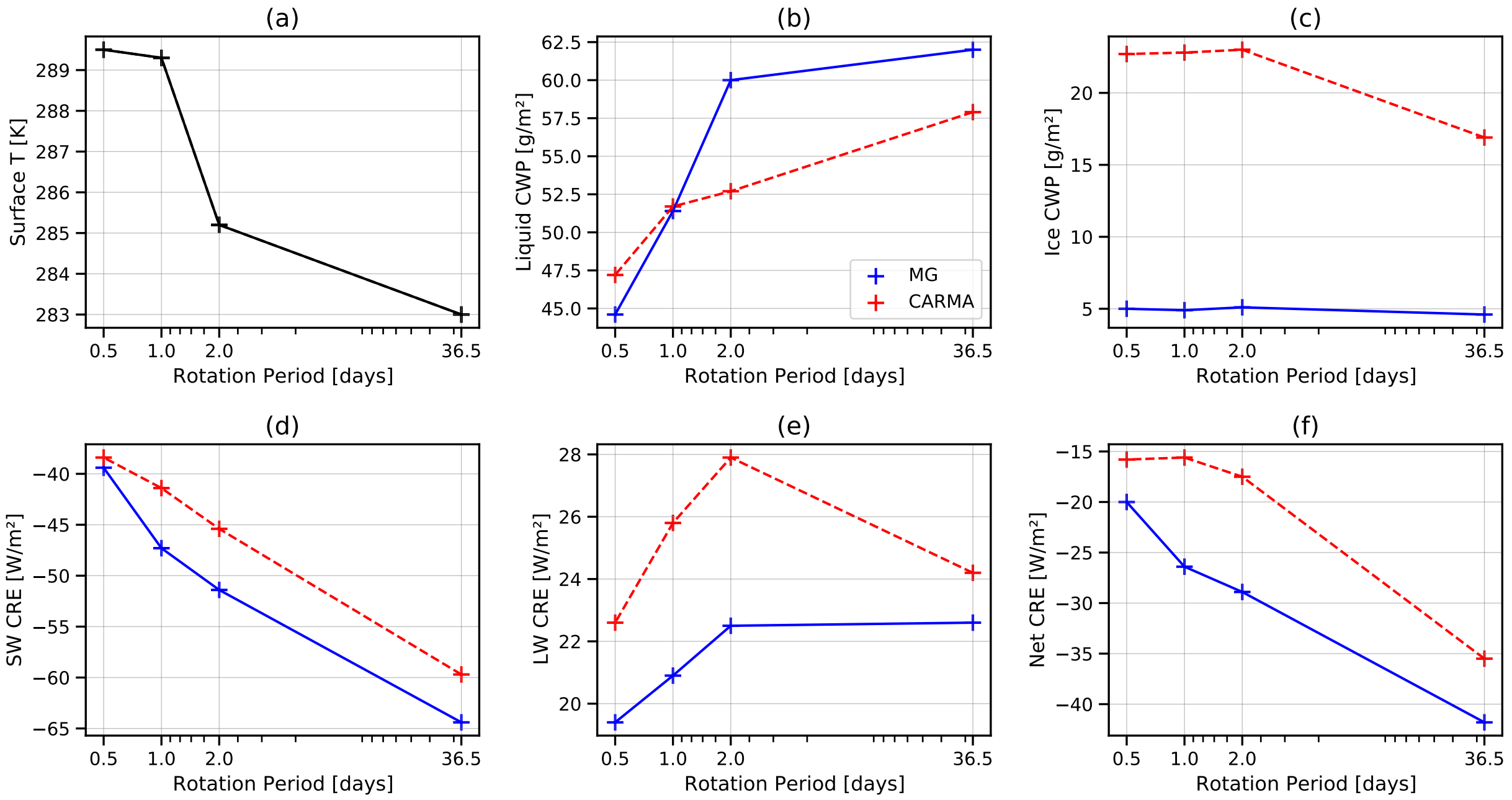

**Figure 2.** Annual and global-mean cloud and radiative properties as a function of planetary rotation period: (a) surface temperature, (b) liquid CWP, (c) ice CWP, (d) shortwave CRE, (e) longwave CRE, and (f) net CRE.

Longwave CRE is generated by the contrast between the surface temperature and the temperature of an optically thick cloud. It is therefore mainly contributed by high-altitude ice clouds in the tropics, which have a low emission temperature. The convective region with high clouds becomes wider as the rotation period increases (Figures 3 and 5). In the fast-rotating cases, with a rotation period less than 2 days, this results in larger longwave CRE (Figure 2(e)). This trend does not extend to the 36.5 day simulation, in which the CARMA longwave CRE decreases and the MG longwave CRE stays similar to the 2 day case (Figure 2(e)). Although the Hadley cell reaches higher latitudes in this case, the convective height is much lower (Figure 5) due to a much lower tropical surface temperature (not shown). A lower cloud height leads to a reduced contrast between the surface and cloud temperature and therefore leads to a reduced longwave CRE. For all rotation periods, CARMA produces a 2–5 $\mathrm{W\,m^{-2}}$ stronger longwave CRE than MG (Figures 2(e) and 3). Both the higher cloud top and the larger ice CWP contribute to this. The ice CWP in CARMA ($\sim 20\ \mathrm{g\,m^{-2}}$) is about 4 times larger than in MG ($\sim 5\ \mathrm{g\,m^{-2}}$; Figures 2(c) and 5). However, the majority of the ice cloud mass produced by CARMA has large particle sizes (Figure 6) and is located at lower altitudes (Figure 5), which leads to a weaker impact on longwave CRE (T. Corti & T. Peter 2009). Therefore, the resulting difference in longwave CRE is not as dramatic as the difference in ice CWP.

## 4. Discussion

Our work can be viewed as a confirmation that a sophisticated parameterized cloud microphysics scheme such as MG can simulate climate variables necessary to ascertain planetary habitability similarly to a bin-resolved scheme such as CARMA when extrapolating to Earth-like worlds with similar assumptions regarding surface and aerosol properties. We found that the difference in net CRE between CARMA and MG microphysics is between 4 and 10 $\mathrm{W\,m^{-2}}$, which is on the order of 1% of absorbed shortwave or emitted longwave radiation. For comparison, changing the rotation period from 0.5 to 36.5 days led to a change in net CRE of 20 $\mathrm{W\,m^{-2}}$ in both models. Rotation period is just one example of a variable that may be unknown in the context of exoplanet climate modeling that likely has a radiative forcing similar to or larger than the difference we found between CARMA and MG microphysics. Other uncertain variables include, for example, continental configuration, atmospheric pressure, greenhouse gas concentrations, and other cloud modeling assumptions (J. Yang et al. 2019). For further context, the radiative forcing of doubling $CO_2$ in modern Earth's climate is 4 $\mathrm{W\,m^{-2}}$, which is too small to affect Earth's habitability (R. M. Ramirez et al. 2014).

Our work is consistent with previous work suggesting that clouds are likely to increase the signal-to-noise ratio (SNR) for observations of biosignatures such as $O_2$ and $O_3$ by HWO (H. Yang et al. 2025). A limitation of H. Yang et al. (2025) is that 1D CARMA only produces low-level stratus and stratocumulus liquid clouds. Our 3D GCM results with both CARMA and MG in this paper show similar albedo increases, suggesting that the main conclusion of our previous work is robust. We confirm this with an example simulation of the detection of $O_2$ by direct imaging using the Planetary Spectrum Generator (G. L. Villanueva et al. 2022) in the same telescope configuration as H. Yang et al. (2025; Figure 7). For both MG and CARMA, the SNR is higher with clouds than without, although the effect is larger with MG, consistent with the higher shortwave CRE simulated by MG. However, in the 0.25–0.5 $\mu$m wavelength range CARMA clouds reflect more strongly than MG clouds, suggesting the observational impact of cloud simulation depends on the absorption bands of biosignature investigated in addition to the cloud scheme.

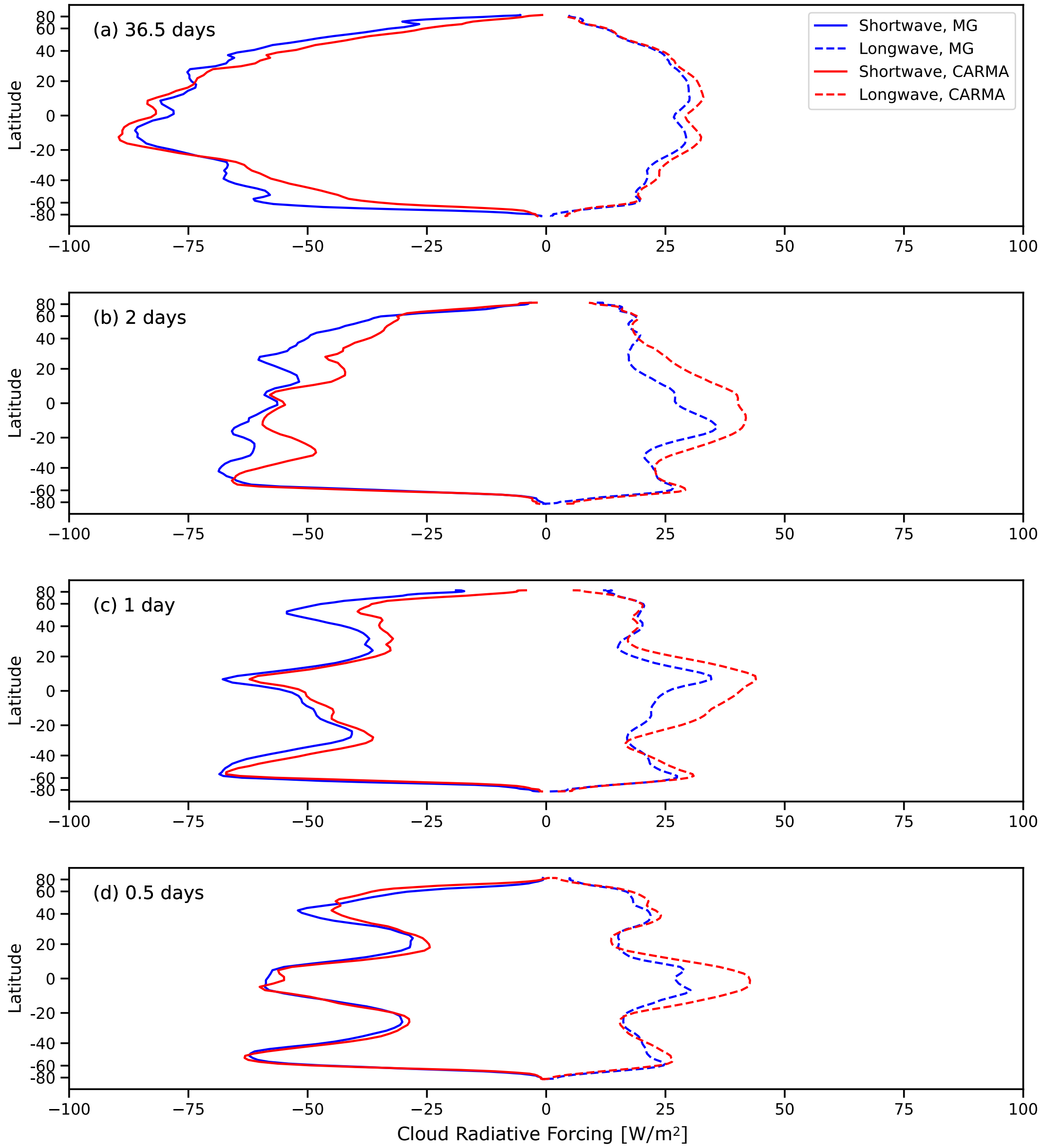

**Figure 3.** Zonal mean longwave (red) and shortwave (blue) CRE for MG (solid) and CARMA (dashed) for different rotation periods (marked in the top left of the panels).

An important implication of our work that deserves follow-up research is that resolved bin cloud microphysics could significantly affect transmission spectral retrievals. This is because CARMA and MG have very different ice cloud size distributions (Figure 6), which are the types of clouds that form at the high altitudes that transmission spectroscopy samples. While MG shows a unimodal size distribution centered around $\sim$70 $\mu$m, CARMA has a bimodal distribution with one mode centered around $\sim$15 $\mu$m and the other centered around $\sim$210 $\mu m$. This difference reflects the increased physicality allowed by detailed process-based bin modeling in CARMA. The small ice particles tend to form through nucleation at high altitudes where both temperature and water vapor content are low. The growth of these particles is strongly controlled by the coagulation process, when the particles collide with each other and form larger particles. The stochastic nature of this process (E. X. Berry & R. L. Reinhardt 1974; D. Lamb & J. Verlinde 2011) will result in a size distribution of cloud particles with two modes, reflecting whether the particles encounter early collisions or not. With a prognostic solution for coagulation, CARMA can better reflect this feature.

Any model can be tuned to roughly reproduce the CRE in modern Earth, but CARMA provides some potential advantages and increased confidence when extrapolating to exoplanets due to the increased physicality of its representation of cloud microphysics. We can see hints of this in detailed comparisons of CARMA with MG for modern Earth, which show that the spatial distribution of liquid clouds simulated by CARMA is closer to the Moderate Resolution Imaging Spectroradiometer observations than MG (C.-C. Liu et al. 2026). This is most clear in the Northern Hemisphere, where CARMA shows an extensive cloudy region over oceans near 40°N, which matches observational results well (C.-C. Liu et al. 2026). In contrast, the corresponding region in MG is shifted to around 50°N–60°N. For ice clouds, it is difficult to determine whether CARMA or MG performs better for modern Earth because observational estimates vary by 1 order of magnitude (K. Meyer et al. 2007; D. I. Duncan & P. Eriksson 2018; W. Wang et al. 2022). That said, C.-C. Liu et al. (2026) show that the large ice cloud water content in CARMA is closer to the ERA5 reanalysis product.

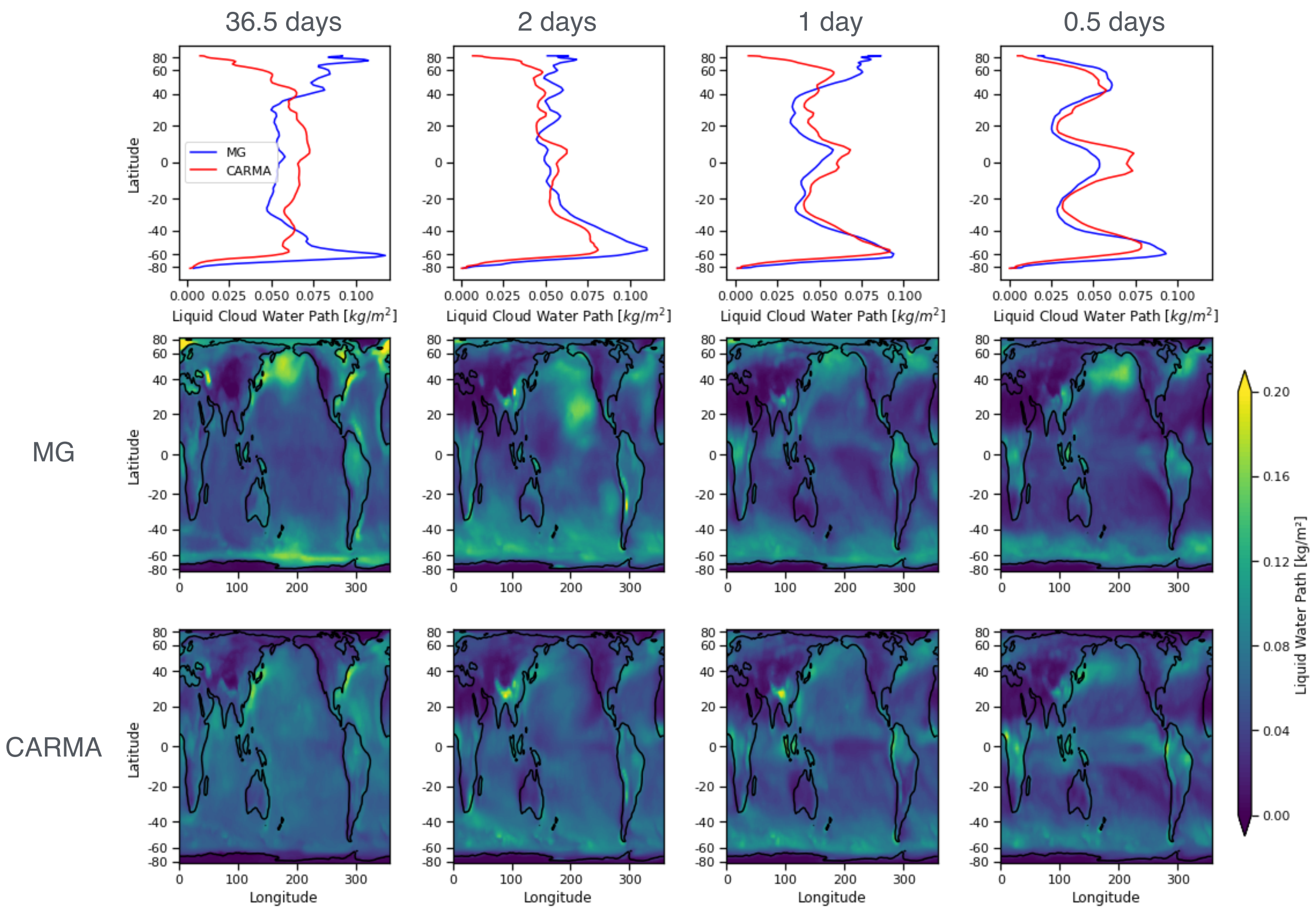

**Figure 4.** Liquid CWP distributions for different rotation periods (columns). The top row shows zonal mean values for MG (blue) and CARMA (red). The middle and bottom rows show annual mean maps for MG and CARMA.

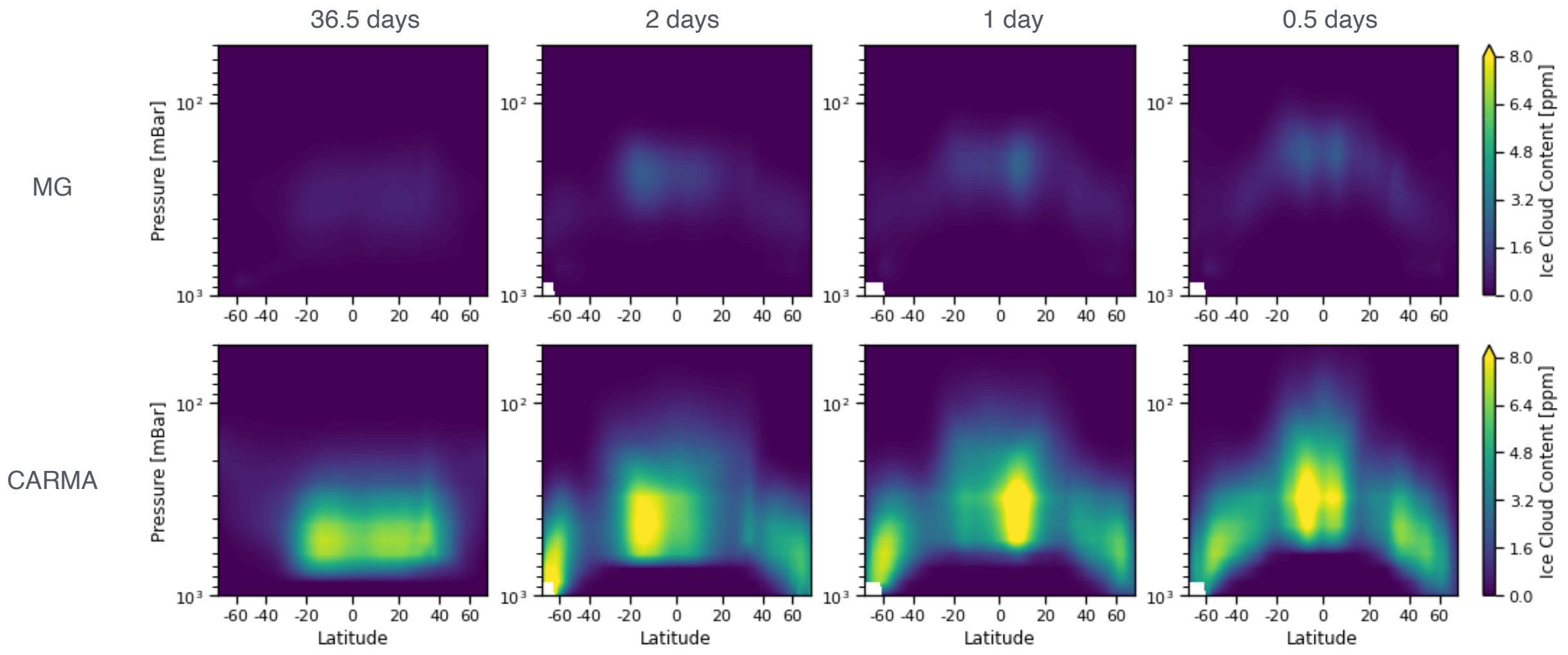

**Figure 5.** Zonal mean ice water content for MG (first row) and CARMA (second row) for different rotation periods (columns).

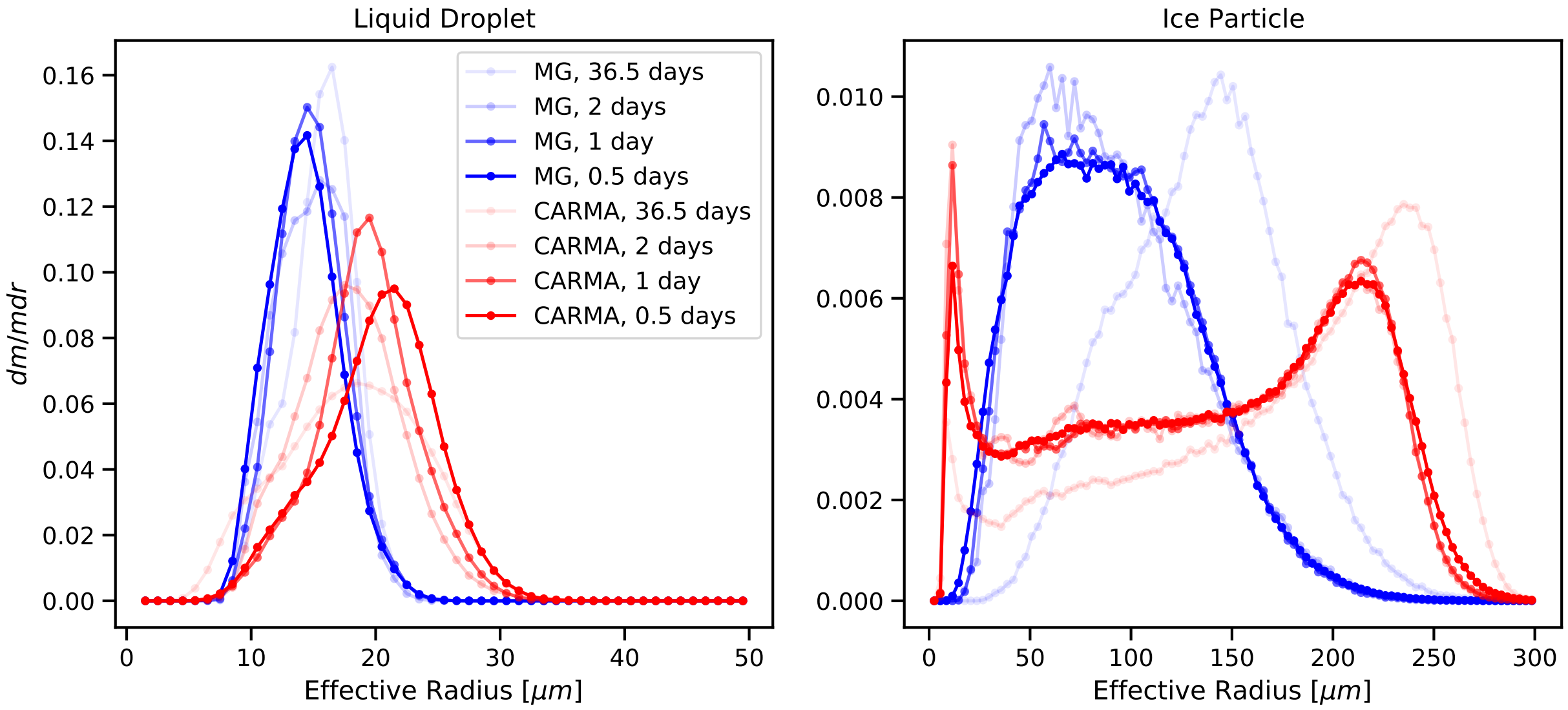

**Figure 6.** Global distribution of effective radius for liquid (left) and ice (right) for MG (blue) and CARMA (red) with transparency indicating rotation periods. Note that the histograms represent the mass of cloud particles.

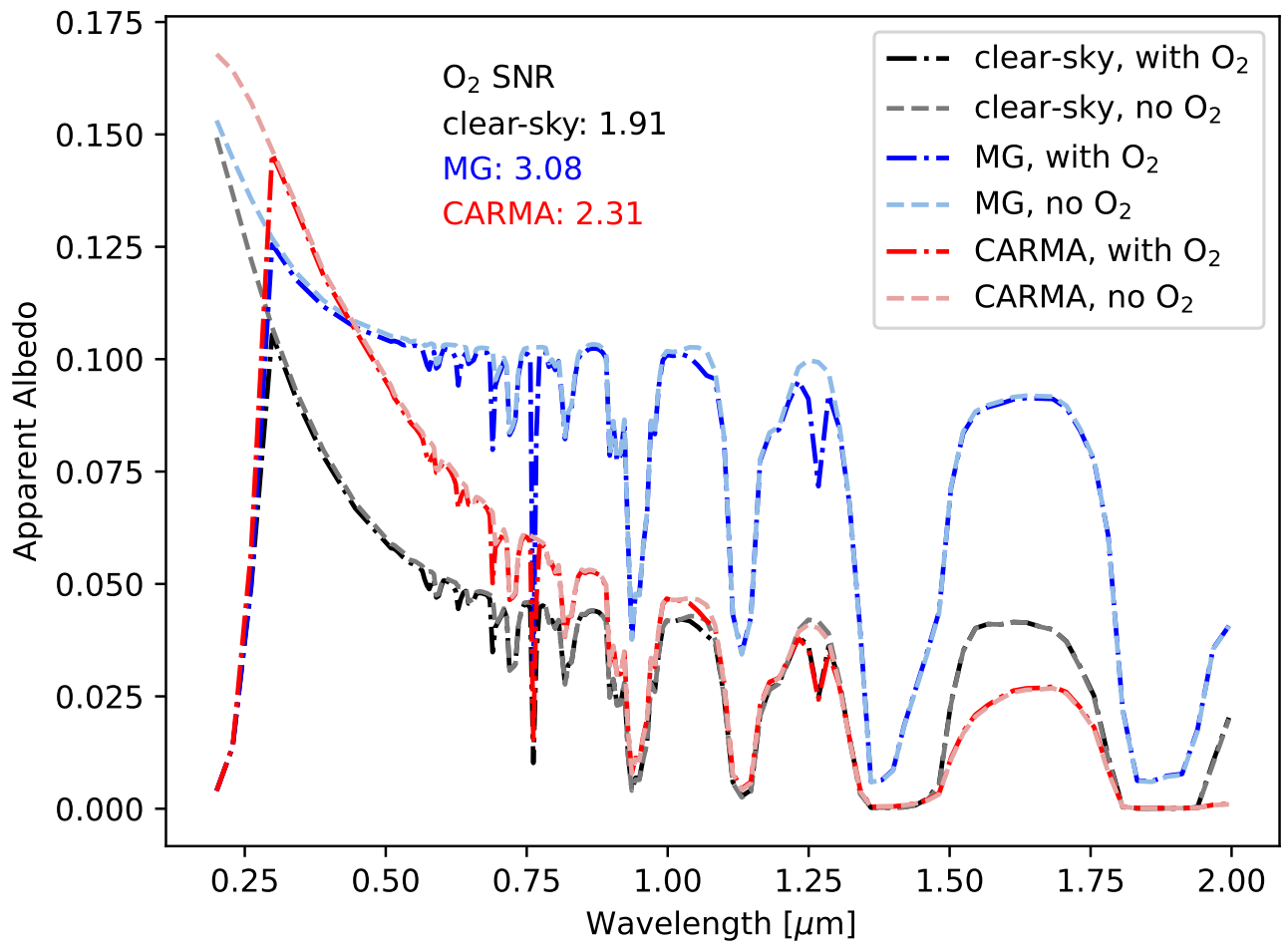

**Figure 7.** Direct imaging spectroscopy simulated with the Planetary Spectrum Generator represented by apparent albedo. The telescope setups and the orbital parameters of the target are the same as in H. Yang et al. (2025), including a phase angle of 90°. The three curves each show an Earth-like planet simulated with CARMA cloud (red), with MG cloud (blue), or cloud free (black). The spectra without $O_2$ in the atmosphere are represented by lighter colors. The SNRs for $O_2$ detection are marked in the figure.

Our work suggests a number of extensions. First, a number of continental configurations and aquaplanet surface boundary conditions could be tested. Second, a dynamic ocean that could calculate changes in ocean heat transport as the rotation rate is varied could be included. Third, CARMA could be coupled to more GCMs, and a model intercomparison could be done, including a comparison with data from other solar system planets. Fourth, the versatile radiative transfer model ExoRT (E. T. Wolf et al. 2022) could be incorporated into CAM6-CARMA so that it could be applied to planets with different stellar spectra and a wider range of atmospheric gases.

## 5. Conclusion

In this work, we studied the impact of the CARMA cloud bin microphysics model embedded in the GCM CESM2/CAM6 on exoplanets with different rotation periods as compared to the standard two-moment MG microphysics scheme. Our main conclusions are as follows:

1. The difference in CRE between CARMA and MG is $\sim$4–10 W $m^{-2}$, which represents a relatively minor forcing on climate and is unlikely to affect the habitability of most exoplanets.
2. CARMA produces a significantly different ice size distribution than MG, which could be an important effect for transmission observations.
3. CARMA produces less negative shortwave CRE than MG, especially in the mid-latitudes, primarily because of lower CWP and larger liquid cloud effective radius.
4. CARMA produces more positive longwave CRE than MG, primarily because its ice cloud water content is larger, which is particularly important for high-altitude clouds.

## Acknowledgments

This work was supported by NASA award No. 80NSSC21K1718, which is part of the Habitable Worlds program. This work was supported by NASA grant No. 80NSSC21K1533, which is part of the Future Investigators in NASA Earth and Space Science and Technology program. Y.Z. has been supported by the National Oceanic and Atmospheric Administration (grant nos. 03-01-07-001, NA17OAR4320101, and NA22OAR4320151).

## ORCID iDs

Huanzhou Yang https://orcid.org/0000-0001-8693-7053
Eric T. Wolf https://orcid.org/0000-0002-7188-1648
Cheng-Cheng Liu (刘成诚) https://orcid.org/0000-0001-9110-6414
Yunqian Zhu https://orcid.org/0000-0001-7751-397X
Owen B. Toon https://orcid.org/0000-0002-1394-3062
Dorian S. Abbot https://orcid.org/0000-0001-8335-6560